\documentclass[12pt]{article}
\usepackage{graphicx}
\usepackage{yf}

\newcommand{\SigmaGreater}{\mathop{\Sigma^{>}}}
\newcommand{\SigmaLess}{\mathop{\Sigma^{<}} }

\newcommand{\GGreater}{G^{>}}
\newcommand{\GLess}{G^{<}}

\newcommand{\ampl}[1]{\mathcal{T}_{#1}}

\newcommand{\Msigma}{\mathop{M_\sigma}}

\newcommand{\crosssection}[1]{\mathop{\sigma(#1)}}

\begin{document}

\title{Pion damping width from\\ $SU(2)\times SU(2)$ NJL model}
%\title{Pion width in hot and dense matter}
\author{D. Blaschke, M. K. Volkov, V. L. Yudichev\\[5mm]
\itshape Joint Institute for Nuclear Research, Dubna}
\date{}
\maketitle
\abstract{ Within the framework of the NJL model, we
investigate the modification of the pion damping width in a hot pion
gas for  temperatures ranging from 0 to 180 MeV. The pion is
found to broaden noticeably at $T >$ 60 MeV. Near the chiral
phase transition $T\sim 180$ MeV, the pion width is saturated
and amounts to 70 MeV. The main contribution to the width comes from
pion--pion collisions. Other contributions are found negligibly small.
}\thispagestyle{empty}

%\large
\clearpage
\section{Introduction}

In the atmosphere of  growing interest in the quark-gluon
plasma, observations of  dilepton (electron--positron pair)
production in relativistic heavy-ion collisions draw more and
more attention of  physicists making efforts to investigate
hadron physics in extreme conditions of temperature
and density. Experimental data have already been taken, and
some recent analyses on the dilepton production in Pb+Au (158
GeV/u) collisions are available, e.~g., from the CERES collaboration
\cite{CERES1,CERES2}.

Insofar as the description of hot and dense media directly
from QCD is not yet
available, the modelling of such processes is an urgent problem.
Recently, an attempt was made to account for the observed
dilepton production rate \cite{CERES1,CERES2},
using the simple Bjorken scenario of the space-time evolution
\cite{Bjorken,Schulze96}
of an ultrarelativistic heavy-ion collision and the vector-dominance model
expression for the pion electromagnetic
form factor \cite{Schulze96}. In \cite{Schulze96}, it was concluded that the
experimental dilepton spectrum could not be explained without assuming
a modification of the $\rho$-meson mass and width in medium.

Besides the $\rho$-meson mass and width modification, the same
is expected for pions. Pions dominate in heavy-ion collisions
and it is very important to study how their properties change
with increasing temperature and density, especially when approaching
a phase transition in a hot matter. The behavior of the
pion in extreme conditions has already been investigated in
models of the NJL type \cite{Asakawa89,Ebert93,Kunihiro94,bbvy2001}
within the mean-field
approximation, which is common in NJL \cite{Volkov86,Ebert94},but
does not take into account  collisions in the pion gas.
However, as we show in our paper, collisions can give a significant
contribution to the pion width in extreme conditions.

For light  particles (like pions) the density increases with
temperature approximately as $T^{3}$. Thus, one expects that the
density is large near the supposed phase-transition temperature
$T_c$, and collisions of particles occur at a much higher rate
than in cold matter. Collisions lead to shorter lifetimes (or
larger widths) of hadronic states at extreme environmental
conditions. In terms of hadron correlators, an additional
imaginary contribution to the hadron self-energy operator comes
from collision integrals \cite{Kadanoff}, thereby broadening all
particles in the hadron gas. The resulting width is closely
related to the process of returning  a disturbed many-particle
system to an equilibrium state: damping. Further, the width thus formed is called
the damping width.

 There are several approaches to calculate the collision integrals. In our work,
we follow the prescription  given by Kadanoff and Baym \cite{Kadanoff}.
This way, the contribution to the hadron self-energy from collision integrals
has a straightforward interpretation: its reciprocal is
the average time between two consecutive collisions, the lifetime. To
estimate the average lifetime of a hadron state,
one needs to know cross sections for different
collision processes averaged over the density of particles
with the Bose amplification and Pauli suppression
taken into account.

One can try to find the modified self-energy of a particle
using a self-consistent functional formalism, as it is described,
e.~g., by Hees and Knoll in \cite{Knoll}. However, in
\cite{Knoll}, the pion damping width was treated  as an external
parameter and was not fixed. Our  purpose is to find an estimate
for the pion damping width from a simple quark model and
investigate possible implications of the  pion broadening. We
calculate cross sections in the framework of the bosonized NJL
model with the infrared cut-off \cite{bbvy2001}. The meson
spectral functions are chosen to be of the Breit--Wigner form.
Then we iterate the equation for the width until a
self-consistent solution is found.

The structure of our paper is as follows. In Section 2, we introduce the
pion damping width. In Section 3, the cross section of the process $\pi\pi\to\pi\pi$
is derived. The numerical results are given in Section~4.
The discussion is given in the last section.

\section{Pion lifetime in hot matter}
In the vacuum, the lifetime of a particular state is determined
by the probability of its decay into other states. Thus, a
stable particle gains no width in the vacuum at all, whereas in
dense medium, a particle can strike  another one and change its
initial state. So the lifetime of a particular state is
determined by its collision rate. There are also inverse
processes that can restore the decayed state due to the interaction of
particles in medium, and then one can find this state again.
This process prolongs the lifetime of the state. Therefore, the
average lifetime of a particular state is determined by the
direct and inverse processes.

The  in-medium hadron properties are usually investigated
in terms of correlators. A two-point correlator  can then be
expressed through a spectral function which  is just
the correlator's imaginary part. For a stable particle,
whose correlator has only one pole, the spectral function
is simply a delta function,
while for a state whose energy is spread, the spectral function
is a continuous function of energy and momentum.

As discussed in \cite{Kadanoff}, the probability, that,
after putting a particle with momentum $p$ to the gas at the time $t$,
removing the particle with  the same momentum at the time $t'$,
one can find the gas at the same state as in
the beginning, decays as $e^{-\Gamma (t' -t)}$, where $\Gamma$
is a constant determining the decay rate and is usually called the width.
If one chooses the spectral function in the Breit--Wigner form
\begin{equation}
A(s)=\frac{M\Gamma}{(s-M^2)^2+M^2\Gamma^2},
\end{equation}
where $M$ is the mass of the state,
the probability indeed decays as $e^{-\Gamma (t' -t)}$ (see \cite{Kadanoff}).
Here, the width is a measure of the dispersion of the state energy.
In general, the width is a function of energy and momentum, and one would
finally come to a set of functional equations for the width,
which are difficult to solve.
In our work, we approximate the pion width by a constant and obtain a simple
equation to be solved by iterations. However, the width is supposed to depend on
the temperature and density.

The pion damping width (or lifetime $\tau$) can be calculated by the following
formula from \cite{Kadanoff}:
\begin{equation}\label{gamma}
\Gamma(p)=\tau^{-1}(p) = \SigmaGreater(p) -\SigmaLess(p),
\end{equation}
where
\begin{equation}\label{sigmaless}
\SigmaLess(p)=\int\limits_{p_1}\int\limits_{p_3}\int\limits_{p_4}
(2\pi)^4\delta_{p_1,p;p_3,p_4}|\ampl{}\, |^2
\GGreater(p_1)\GLess(p_3)\GLess(p_4),
\end{equation}
\begin{equation}\label{sigmagreater}
\SigmaGreater(p)=\int\limits_{p_1}\int\limits_{p_3}\int\limits_{p_4}
(2\pi)^4\delta_{p_1,p;p_3,p_4}|\ampl{}\, |^2
\GLess(p_1)\GGreater(p_3)\GGreater(p_4),
\end{equation}
in accordance with the notation given in \cite{Kadanoff}. Here,
$\ampl{}$ is the process amplitude (will be given in Section 3),
$\GGreater_{i}(p)\!=\![1+n_i(\vec{p},s_i,T)] A_i(p^2)$;\quad
$\GLess_{i}(p)=n_i(\vec{p},s,T)A_i(p^2)$, with $n_i$ being the
boson occupation numbers
\begin{equation}
n_i(\vec{p},s_i,T)=\left[\exp\left(\frac{\sqrt{\vec{p}^2+s_i}}{T}\right)-1\right]^{-1}
\end{equation}
and $A_i(p^2)$ the spectral function of the $i$th state.
We also use the notations
$\int\limits_{p_i}=\int\frac{d^4 p_i}{(2\pi)^4} $ and
$\delta_{p_1,p_2;p_3,p_4}=\delta(p_1+p_2-p_3-p_4)$ in (\ref{sigmaless}) and (\ref{sigmagreater}),
omitting the subscript at $p_2$ to comply with the general definition of
width (\ref{gamma}); $p_2=p$ is assumed throughout the rest of the paper.
The integration is performed in the four-dimensional momentum space
over the momenta $p_1$, $p_3$, $p_4$.  The indices 1 and 2 correspond
to the initial states, while 3 and 4 to the final ones. The width $\Gamma$
is calculated for pion 2 rested in the heat-bath frame.

For the inverse lifetime of a pion, one thus obtains
\begin{eqnarray}\label{Gamma}
&&\tau^{-1} =\Gamma=
\int\frac{d^3\vec{p}_1}{(2\pi)^3}\int d s_1 v_{\rm
rel} A_\pi(s_1)\times\nonumber\\
&&\quad\times[n_\pi(\vec{p}_1,s_1,T)  {\sigma^{\rm dir}}^\ast(s;s_1,s_2)-\nonumber\\
&&\quad-(1+ n_\pi(\vec{p}_1,s_1,T))  {\sigma^{\rm inv}}^\ast(s;s_1,s_2)],
\end{eqnarray}
where $v_{\rm rel}$ is the relative velocity of particles 1 and 2, and
${\sigma^\mathrm{dir}}^\ast(s;s_1,s_2)$ and ${\sigma^\mathrm{inv}}^\ast(s;s_1,s_2)$
are the averaged cross sections for the ``direct'' and ``inverse'' processes,
respectively, with the probability of the final
states to be off-mass-sehll  taken into account:
\begin{equation}\label{crosssection}
{\sigma^\mathrm{dir(inv)}}^\ast(s;s_1,s_2)=\int d s_3\int d s_4 A_\pi(s_3)
A_\pi(s_4) \sigma^\mathrm{dir(inv)}(s;s_1,s_2,s_3,s_4).
\end{equation}
(See (\ref{s}) and (\ref{si}) for definitions of $s$ and $s_i$.)

What one needs then is the cross sections of the processes under investigation.
The amplitudes of these processes can be calculated in an
effective model of pion interaction.  In the next section,
we calculate the amplitudes and cross sections in the framework
of the NJL model with the infrared cut-off
\cite{bbvy2001}.

\section{Pion--pion scattering amplitude and cross section from NJL model}

\subsection{The $\pi\pi\to \pi\pi$ amplitude}
Let us consider  scattering of $\pi^0$  on a pion from the medium:
 $\pi^0\pi^0\to$ $\to\pi^0\pi^0$,  $\pi^0\pi^0\to\pi^+\pi^-$,
 $\pi^0\pi^\pm\to\pi^0\pi^\pm$.
 We allow also for the lightest scalar isoscalar resonance, the
$\sigma$-meson, as  an intermediate state
($\pi\pi\to\sigma\to\pi\pi$), because of its importance shown in
various investigations of the pion--pion interaction
\cite{Kunihiro94,Volkov86,Volkov02,Huefner95}. Here, we use an
$SU(2)\times SU(2)$ chiral quark model of the NJL type
\cite{Volkov86,Ebert94} where, using the bosonization
procedure, one obtains  the Lagrangian for pions and
$\sigma$-mesons. The part that contains three- and four-particle
vertices has the form
\begin{eqnarray}
&&\mathcal{L}_{\rm int}=2 m g_\sigma\sigma^3
+ 2 m g_\pi \sqrt{Z} \sigma (2\pi^+\pi^- + (\pi^0)^2) -g_\pi^2\sigma^2\pi^2-
\nonumber\\
&&\qquad-\frac{g_\pi^2 Z}{2} (4\pi^+\pi^-\pi^+\pi^-
+ 4 \pi^+\pi^-(\pi^0)^2 + (\pi^0)^4) -\frac{g_\sigma^2}{2}\sigma^4.
\end{eqnarray}
Here, $m$ is the constituent quark mass ($m=g_\pi f_\pi$;
$f_\pi=93$ MeV is the pion weak decay constant in the vacuum).
The constants $g_\pi$ and $g_\sigma$ describe the interaction
of the pion and $\sigma$-meson with quarks, respectively. They
are related to each other by the equation
\begin{equation}
g_\pi=g_\sigma \sqrt{Z}.
\end{equation}
The constant $Z$ originates from $\pi$--$a_1$ transitions and is
equal to
\begin{equation}
Z=\left(1-\frac{6m^2}{M_{a_1}^2}\right)^{-1},
\end{equation}
where $M_{a_1}^2=1250$ MeV is the mass of $a_1$-meson. The
values of the constituent quark mass and the constants $g_\pi$
and $g_\sigma$ were calculated in \cite{bbvy2001}. In the vacuum
we have: $m=242$ MeV, $g_\pi=2.61$, and $g_\sigma=2.18$. Their
values at finite temperatures and zero chemical potential were
calculated in \cite{bbvy2001}; here we use these results as
input data in our calculations.

In our approach, the total amplitude of $\pi^0\pi^0\to\pi^0\pi^0$ consists of
a contact term and three resonant contributions in the scalar channel
(see Figs.~1 \textit{a---d})
\begin{equation}
\ampl{\pi^0\pi^0\to\pi^0\pi^0}= -24g_\pi^2
+ 3\ampl{\sigma}(s) +3\ampl{\sigma}(t)+3\ampl{\sigma}(u),
\end{equation}
where
\begin{equation}
\ampl{\sigma}(x) = \frac{4 g_\pi^2 m^2}{ \Msigma^2 - x - i \Msigma\Gamma_{\sigma}},
\end{equation}
and $s, t, u$ are kinematic invariants:
\begin{eqnarray}
s&=&(p_1+p_2)^2=(p_3+p_4)^2,\label{s} \\
t&=&(p_1-p_3)^2=(p_2-p_4)^2,\\
u&=&(p_1-p_4)^2=(p_2-p_3)^2,\label{stu}
\end{eqnarray}
for which the following identity is satisfied:
\begin{equation}\label{si}
s+t+u=s_1+s_2+s_3+s_4; \quad s_i=p_i^2.
\end{equation}

The momentum $p_1$ corresponds to the impacting pion, while
$p_2$ relates to the pion rested in the heat bath. The momenta
$p_3$ and $p_4$ are those for the particles  produced after
collision.

For the charged pions in the final state,  the amplitudes
contain two terms:
\begin{equation}
\ampl{\pi^0\pi^0\to\pi^+\pi^-}= -8g_\pi^2
+ 2\ampl{\sigma}(s)
\end{equation}
for the process $\pi^0\pi^0\to\pi^+\pi^-$ (Figs.~1\textit{a} and 1\textit{b});
\begin{equation}
\ampl{\pi^0\pi^\pm\to\pi^0\pi^\pm}= -8g_\pi^2
+ 2\ampl{\sigma}(t)
\end{equation}
for $\pi^0\pi^\pm\to\pi^0\pi^\pm$ (Figs.~1\textit{a} and 1\textit{c}).

In medium,  model parameters depend on temperature and density.
In our work, as only mesons are considered, we restrict
ourselves to the case of zero chemical potential for which
the temperature dependence of the model parameter has been obtained
in \cite{bbvy2001}.  We use these results in \cite{bbvy2001}  as
input data in our calculations.

\subsection{Cross sections}

The differential cross section for a pion--pion scattering is
determined by the equation
\begin{equation}\label{sigmadir}
\frac{d \sigma^{\rm dir}}{d t}=\frac{|\ampl{} |^2 (1+ n_3)(1+
n_4)}{64\pi s |\vec{p}_{1, {\rm cm}}|^2}
\end{equation}
for the direct and
\begin{equation}\label{sigmainv}
\frac{d \sigma^{\rm inv}}{d t}=\frac{|\ampl{} |^2  n_3
n_4}{64\pi s |\vec{p}_{1, {\rm cm}}|^2}
\end{equation}
for the inverse processes. Here,  $\vec{p}_{\rm 1, cm}$ is the momentum
of pion 1 in the center-of-mass frame for  pions 1 and 2;
$n_3$ and $n_4$ are the occupation numbers of produced
pions.

Having integrated over $t$ in the interval defined by the
lower and upper limits
\begin{eqnarray}
t_{\rm min}&=&
    \left(\frac{s_1-s_2-s_3+s_4}{2\sqrt{s}}\right)^2-(|\vec{p}_{1,\rm
cm}|+|\vec{p}_{3,\rm cm}|)^2,\\
t_{\rm max}&=&
    \left(\frac{s_1-s_2-s_3+s_4}{2\sqrt{s}}\right)^2-(|\vec{p}_{1,\rm
cm}|-|\vec{p}_{3,\rm cm}|)^2,
\end{eqnarray}
we obtain the total cross section.
The center-of-mass momenta of pions  are determined by the equation
\begin{equation}
|\vec{p}_{i,\rm cm}|=\sqrt{E_{i,\rm cm}^2 - s_i}, \qquad i=1,3,
\end{equation}
where the energies $E_{i,\rm cm}$ are defined in the center of mass frame
\begin{eqnarray}
E_{1, \rm cm}&=&\frac{s + s_1 - s_2}{2\sqrt{s}},\\
E_{3, \rm cm}&=&\frac{s + s_3 - s_4}{2\sqrt{s}}.
\end{eqnarray}
To calculate the occupation numbers $n_i$ in (\ref{sigmadir}) and (\ref{sigmainv}),
one needs to know the energies of pion 3 and pion 4 in the heat-bath frame
\begin{equation}
E_3=\frac{s_2 + s_3 -t}{2\sqrt{s_2}},\qquad E_4=\frac{s_2 + s_4
-t}{2\sqrt{s_2}}.
\end{equation}

In the case of pion--pion collisions, four processes contribute to
the total cross section:
\begin{eqnarray}\label{pipi2pipi:crosssection}
&&\crosssection{\pi\pi\to \pi\pi}= \crosssection{\pi^0\pi^0\to\pi^0 \pi^0}+
\crosssection{\pi^0\pi^0\to\pi^+ \pi^-}+\nonumber\\
&&\quad+2\crosssection{\pi^0\pi^+\to\pi^0\pi^+}.
\end{eqnarray}
The process $\pi^0\pi^+\to\pi^0\pi^+$ occurs at the same rate as
$\pi^0\pi^-\to\pi^0\pi^-$, so we do not calculate them
separately; we just put the factor 2 at the last term in
(\ref{pipi2pipi:crosssection}). To obtain the pion damping width,
we substitute the obtained cross sections into Eqns.~(\ref{Gamma}) and (\ref{crosssection}).

\section{Numerical results}
Using the definition of the pion lifetime  given in Section~2 and
the cross sections determined in Section~3, we evaluate
numerically the pion damping width at the temperatures ranging
from 0 to 180 MeV. The upper limit on the temperature scale
corresponds to the expected  transition from the phase with
broken chiral symmetry to the symmetric phase. The resulting
curves are shown on Fig.~2.

All calculations are performed for a neutral pion rested in the
heat bath frame. The pion self-energy is approximated by a constant.

As one can see from Fig.~2, the pion state broadens
noticeably in a hot matter, as compared to the vacuum state,
already at $T\approx 60$ MeV. At $T=160$, the pion--pion
scattering accounts for about 80 MeV in the total width; this is the
maximum value. Then the curve in Fig.~2 turns down; this
behavior of the damping width is caused by a noticeable
decrease of the constant $g_\pi$ after $T=160$ MeV.
The cross section is proportional to $g_\pi^4$, whose value
reduces by half at $T=180$ MeV, comparing to $T=160$ MeV. For
the temperatures from 160 MeV to 180 MeV, this weakening of the
pion--pion interaction overpowers the expected increase in
collision integrals.

Besides the pion--pion scattering, we estimated also some other
possible contributions to the pion width. They come from the
following processes: $\pi\pi\to\sigma\sigma$,
$\pi\sigma\to\pi\sigma$, $\pi\pi\to\sigma$, and $\pi\pi\to \bar
qq$. All these contributions were calculated separately, i.~e.,
with other modes switched off. (This gives the upper limit for a
solution of the equation for width, because the collision
integrals decrease if the spectral functions broaden.) They turned
out to be small. At $T=180$ MeV (which is a little below
$T_c=186$ MeV \cite{bbvy2001}), $\pi\sigma\to\pi\sigma$ gives
about 5 MeV, $\pi\pi\to\sigma\sigma$ about 1 MeV, and
$\pi\pi\to\sigma$ even less than 1 MeV. The decays to quarks  contribute
to the pion width less than 7 MeV. The smallness of these contributions
was the reason to discard them in our calculations whose purpose is
to make a qualitative estimate for the pion damping width.

\section{Discussion and conclusion}
In the framework of the $SU(2)\times SU(2)$ NJL model, the pion
damping width was calculated for the range of temperatures from
0 to 180 MeV. The definition for the damping
width, given in \cite{Kadanoff}, was used. A self-consistent
method where the pion width does not depend on energy and momentum
was used to estimate the contributions to the pion damping width
from  pion--pion scattering in a hot and dense matter.
Upper limits for
alternative contributions were estimated qualitatively.
It was found that  the pion--pion scattering is dominant,
and other processes that give small contributions can be discarded.

In our investigation we came to the conclusion that, in a hot
gas, the pion spectral function significantly broadens, while
nearing the phase transition point. This may have many
implications for various processes in a hot and dense matter
where pions are involved.
In particular, this can affect the dilepton production through pion--pion annihilation
in heavy-ion collisions \cite{Schulze96}.

Of course, a more systematic and formal approach, like  one
suggested by Hees and Knoll \cite{Knoll}, would be more
preferable for a study of processes occurring in a hot matter.
Nevertheless, our approach, which stems from Quantum Statistical
Mechanics, has a clear physical interpretation, and it allowed us
to make a prediction for the behavior of the pion damping width in
a hot matter. Furthermore, similar calculations can be done for
other particles, e.~g., for the $\sigma$- and $\rho$-mesons.
The $\sigma$-meson width   at those
temperatures, at which the direct decay $\sigma\to\pi\pi$ is
suppressed and contributions from collision integrals to the
$\sigma$-meson width become noticeable, is of particular interest.
Moreover, the diagrams with
the $\sigma$-pole play an important role in different processes
occurring in a hot matter.

This work has been supported by RFBR Grant no.~02-02-16194 and Heisen\-berg--Landau program.

\clearpage
\section*{Figure Captions}
\begin{enumerate}
\item[Figure 1.] Diagrams contributing to the  $\pi\pi\to\pi\pi$ amplitude.
\item[Figure 2.] Pion damping width as a function of $T$.
\end{enumerate}

\clearpage
\section*{Figures}

\begin{figure}[h]
\label{diag}
\caption{Diagrams contributing to the  $\pi\pi\to\pi\pi$ amplitude.}
\vspace{1cm}
\begin{center}
\includegraphics[scale=0.9]{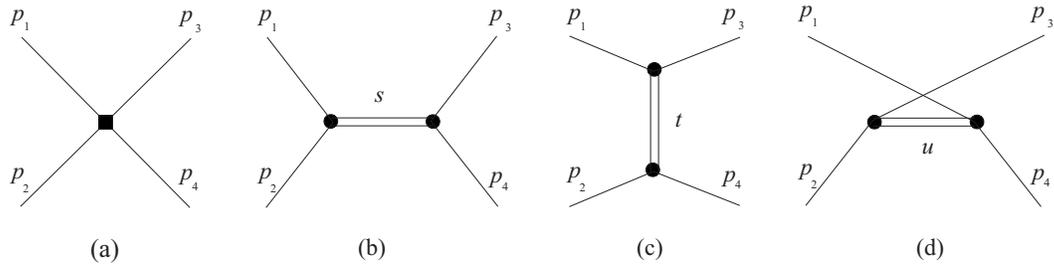}
\end{center}
\end{figure}

\clearpage
\begin{figure}[h]
\label{pdw}
\caption{Pion damping width as a  function of $T$.}
\vspace{1cm}
\begin{center}
\includegraphics[scale=0.8]{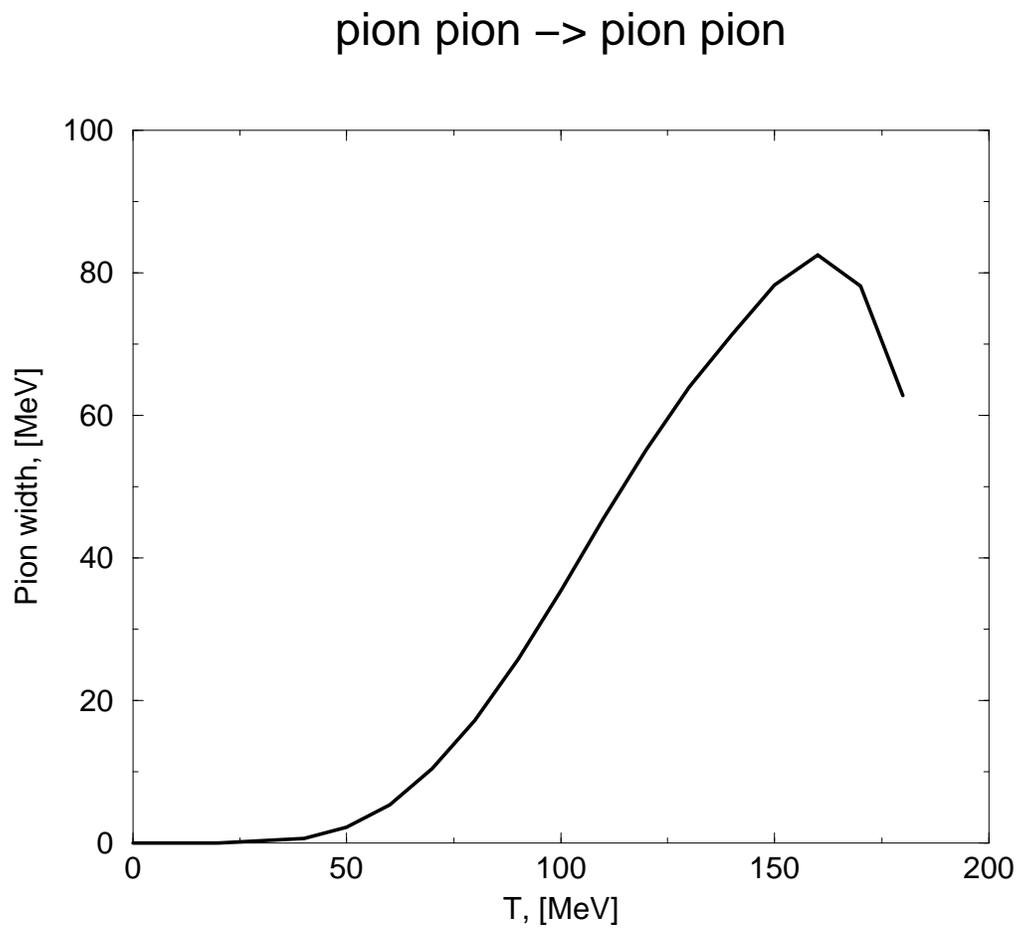}
\end{center}
\end{figure}

\end{document}